\documentclass[aps,prl,nofootinbib,superscriptaddress,twocolumn]{revtex4}
\usepackage{graphicx}
\usepackage{color}
\usepackage[normalem]{ulem}

\begin{document}

\title{Direct spectroscopic observation of a shallow hydrogen-like donor state in insulating SrTiO$_{3}$}
\author{Z.~Salman}
\email{zaher.salman@psi.ch}
\author{T.~Prokscha}
\email{thomas.prokscha@psi.ch}
\author{A.~Amato}
\author{E.~Morenzoni}
\author{R.~Scheuermann}
\author{K.~Sedlak}
\author{A.~Suter}
\affiliation{Paul Scherrer Institute, Laboratory for Muon Spin Spectroscopy, CH-5232 Villigen PSI, Switzerland}

\date{\today}

\begin{abstract}
  We present a direct spectroscopic observation of a shallow
  hydrogen-like muonium state in SrTiO$_3$ which confirms the
  theoretical prediction that interstitial hydrogen may act as a
  shallow donor in this material. The formation of this muonium state
  is temperature dependent and appears below $\sim 70$~K. From the
  temperature dependence we estimate an activation energy of $\sim
  50$~meV in the bulk and $\sim 23$~meV near the free surface. The
  field and directional dependence of the muonium precession
  frequencies further supports the shallow impurity state with a rare
  example of a fully anisotropic hyperfine tensor. From these
  measurements we determine the strength of the hyperfine interaction
  and propose that the muon occupies an interstitial site near the
  face of the oxygen octahedron in SrTiO$_3$. The observed shallow
  donor state provides new insight for tailoring the electronic and
  optical properties of SrTiO$_{3}$-based oxide interface systems.
\end{abstract}
\maketitle

The discovery of a high mobility two-dimensional electron gas (2DEG)
at the interface between two insulating Perovskite oxides;
TiO$_{2}$-terminated SrTiO$_{3}$ (STO) and LaAlO$_{3}$
(LAO)\cite{Ohtomo04N,Thiel06S,Huijben06NM} has prompted great interest
in these oxides. In addition to the 2DEG, this interface was found to
be magnetic\cite{Brinkman07NM,BenShalom09PRB,Salman12PRL} and even
superconducting below $\sim300$~mK\cite{Reyren07S}. It is generally
agreed that the high carrier densities at the interface are associated
with various effects, including doping with electrons or oxygen
vacancies \cite{Thiel06S, Pentcheva06PRB, Park06PRB, Takizawa06PRL,
  Kalabukhov07PRB}, inter-diffusion \cite{Takizawa06PRL, Nakagawa06NM,
  Willmott07PRL}, and the influence of lattice distortions
\cite{Ahn03N, Gemming06AM, Hamann06PRB, Maurice06PSS, Okamoto06PRL}.
These discoveries provide an interesting prospect for producing
interfaces with physical properties not present, nor predictable, from
the constituent materials, and may lead to new technological
applications. Here we discuss the response of these oxides to
impurities, in particular hydrogen, which may play an important role
in the discovered phenomena. Moreover, we note that the reported
results are extremely relevant for possible applications of STO and
related insulating oxides in fuel cells and hydrogen sensors
\cite{Higuchi01PRB,Higuchi98PRB,Yukawa99SSI}.

Hydrogen is an ubiquitous impurity in device or sample fabrication and
can cause unintended modifications of the electronic and structural
properties. For example, the properties of the 2DEG at STO/LAO
interfaces may be affected if additional free charge carriers are
present due to unintentional doping \cite{Son10JPCM}. The electronic
behaviour of interstitial hydrogen can be characterized by the
position of the H(+/-) level in the band gap, where the formation
energies of H$^+$ and H$^-$ are equal. If this level is close to or
intersecting the conduction/valence band, hydrogen will act as a
shallow donor/acceptor. In elemental and binary semiconductors, a
universal alignment of the H(+/-) level at $\sim$ -4.5~eV with respect
to the vacuum level has been found theoretically \cite{vandeWalle03N}
and supported by experiments \cite{Lichti08PRL}.  The model predicted
successfully hydrogen shallow donor states in ZnO and InN
\cite{Cox01PRL,Hofmann02PRL,Davis03APL}. In this model the H(+/-)
level coincides approximately with the host's charge neutrality level
(CNL). In general the CNL is found to be located at relatively
constant energies with respect to the vacuum level
\cite{vandeWalle03N,Xiong07JAP}.  The coincidence of the H(+/-) level
with the CNL can be understood as follows. In a binary semiconductor
H$^+$ and H$^-$ behave similarly, i.e. they break a bond at the anion
or cation site, respectively, leaving a dangling bond at the opposite
cation or anion site. The H(+/-) level is then located midway in
between the energy levels of the dangling bonds, which corresponds to
the CNL.

In oxides the situation is different, since H$^+$ tends to form an
OH$^-$ antibonding state, without breaking a cation--O bond. H$^-$, on
the other hand, causes the formation of an oxygen dangling bond by
breaking a metal--O bond. Thus, the H(+/-) level is determined by the
average of the oxygen dangling bond and OH$^-$ antibonding levels,
which is located at relatively constant energies of about $4.5-5.5$~eV
above the valence band maximum
\cite{Xiong07JAP,Peacock03APL}. Therefore, hydrogen should form a
shallow donor state in oxides with a band gap of less than $\sim
4.5$~eV, such as STO which has an indirect band gap of $\sim
3.2$~eV. This prediction has been supported recently by a new
theoretical work on hydrogenated vacancies and hidden hydrogen in STO
\cite{Varley14PRB}. In contrast, the universal alignment model
described above places H(+/-) deep in the band gap, i.e. $\sim 0.5$~eV
below the conduction band minimum, which equals the electron affinity
of STO and is located at about $4.0$~eV below the vacuum level. Note,
the model in Refs.~\cite{Xiong07JAP,Peacock03APL} predicts shallow
donor states in SnO$_2$, TiO$_2$, ZrO$_2$, and HfO$_2$ which have been
observed or inferred in muon spin rotation ($\mu$SR) experiments
\cite{Cox06JPCM2}. In these experiments, positive muons, when
implanted into insulators or semiconductors, can capture an electron
to form interstitial muonium [Mu=($\mu^+\rm{e}^-$)], which can be
considered as a light hydrogen isotope and mimics its chemical and
electrical interactions. In fact, a large amount of information on the
structure and electrical activity of isolated interstitial H states in
semiconductors and insulators has been obtained by Mu spectroscopy
\cite{Patterson88RMP,muSR2010}.

In this Letter we present a direct spectroscopic observation of a
shallow hydrogen-like Mu state in a STO (100) single crystal, in
agreement with the model suggested by
Refs.~\cite{Xiong07JAP,Peacock03APL}. Our results are consistent with
muons occupying an interstitial site in the lattice, between two O--O
bonds near the face of the oxygen octahedron. We find that up to $\sim
60$\% of the implanted muons form a shallow muonium state at 25~K with
a relatively small hyperfine coupling. From the field dependence of
the Mu characteristic precession frequencies we find that the
hyperfine tensor is fully anisotropic and estimate the hyperfine
interaction along the principal axes of the tensor $A_X=1.4 \pm 0.1$
MHz, $A_Y=6.7\pm 0.1$ MHz and $A_Z=11.5 \pm 0.1$ MHz. This Mu state is
one of the rare clear examples of a fully anisotropic Mu.

The $\mu$SR experiments were performed on the DOLLY, GPS and LEM
spectrometers at the Paul Scherrer Institut in Villigen, Switzerland.
In these experiments $100 \%$ polarized positive muons are implanted
into the sample. Each implanted muon decays (lifetime $\tau_{\mu}=2.2$
$\mu$s) emitting a positron preferentially in the direction of its
polarization at the time of decay. Using appropriately positioned
detectors, one measures the asymmetry of the muon beta decay along
different directions as a function of time, $A(t)$, which is
proportional to the time evolution of the muon spin polarization.
$A(t)$ depends on the electronic environment of the muon and is used
to extract information on the hyperfine interaction between the muon
and the electrons in the system. Muonium is spectroscopically
identified by its characteristic precession frequencies which allow to
determine the Mu hyperfine parameters \cite{muSR2010,Patterson88RMP}.
In a low energy $\mu$SR (LE-$\mu$SR) experiment, the energy of the
implanted muons, $E$, can be tuned (1-30 keV) to perform a measurement
of $A(t)$ at depths in the range $\sim 1-200$ nm
\cite{Morenzoni03PB,Prokscha08NIMA}. The measurements reported here
were performed on a $15 \times 15 \times 1$~mm single side epitaxially
polished $(100)$ STO single crystal substrate (Crystal GmbH). In the
bulk $\mu$SR measurements, the sample was suspended on an aluminized
Mylar tape and mounted into a He gas flow cryostat. The muons were
implanted with their polarization nominally along $\langle 100
\rangle$ with the field applied perpendicular to it (nominally along
$\langle 010 \rangle$). In the LE-$\mu$SR measurements the sample was
glued to a cold finger cryostat, with the field applied along $\langle
100 \rangle$ and the polarization of implanted muons perpendicular to
it.

\begin{figure}[ht]
  \centering
  \includegraphics[width=0.8\columnwidth]{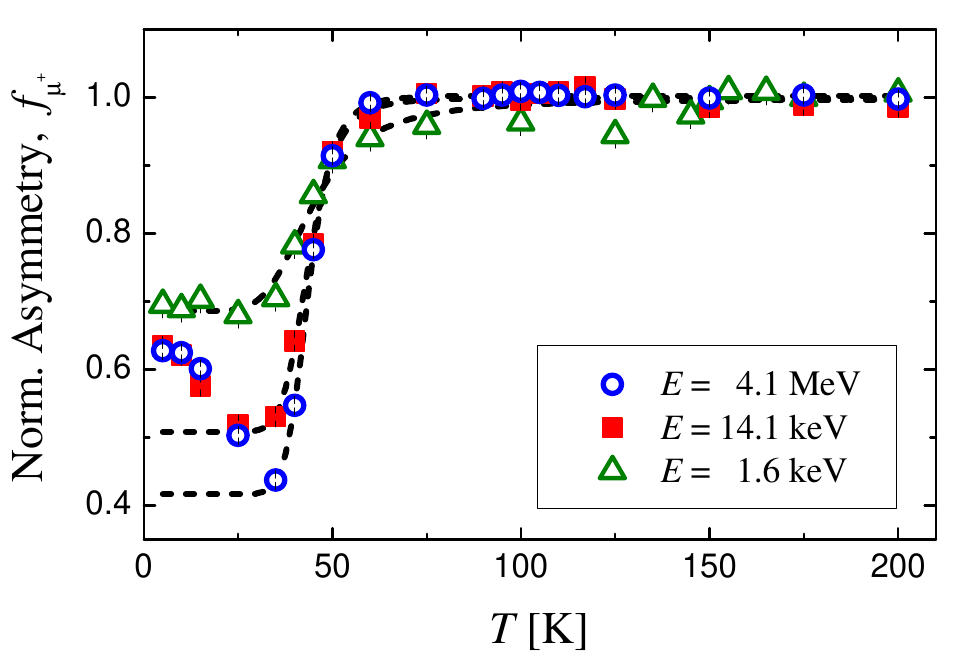}
  \caption{(Color online) The normalized diamagnetic fraction of the
    $\mu^+$ precession signal at $B=10$~mT as a function of
    temperature.  Circles, squares and triangles are measurements at
    4.1 MeV ($B \perp \langle100\rangle$), 14.1 keV and 1.6 keV
    ($B \parallel \langle100\rangle$), respectively. The drop in
    asymmetry below $\sim 70$ K is due to formation of muonium. The
    dashed lines are fits to Eq.~\ref{Activation}.}
  \label{AsyT}
\end{figure}
We start by looking at muon spin precession measurements in a field of
$B=10$~mT applied perpendicular to the muon's initial spin
direction. At room temperature we find that all muons implanted at 4.1
MeV in bulk STO precess at the Larmor frequency of $\mu^+$,
$\nu_0=\gamma_\mu/2\pi~B$, with almost no damping. Here,
$\gamma_\mu/2\pi = 135.5$~MHz/T is the muon gyromagnetic ratio. As we
decrease the temperature, we find that below $\sim 70$ K the amplitude
of the signal precessing at $\nu_0$ (the so called diamagnetic
fraction, $f_{\mu^+}$) decreases sharply, reaches a minimum at $\sim
30$ K, and then increase at lower temperatures (Fig.~\ref{AsyT}).
Similar behaviour is observed at lower muon implantation
energies. However, in this case the decrease in the $f_{\mu^+}$ is
smaller, and at $E=1.6$ keV we observe no increase at low
temperatures.

\begin{figure}[ht]
  \centering
  \includegraphics[width=0.8\columnwidth]{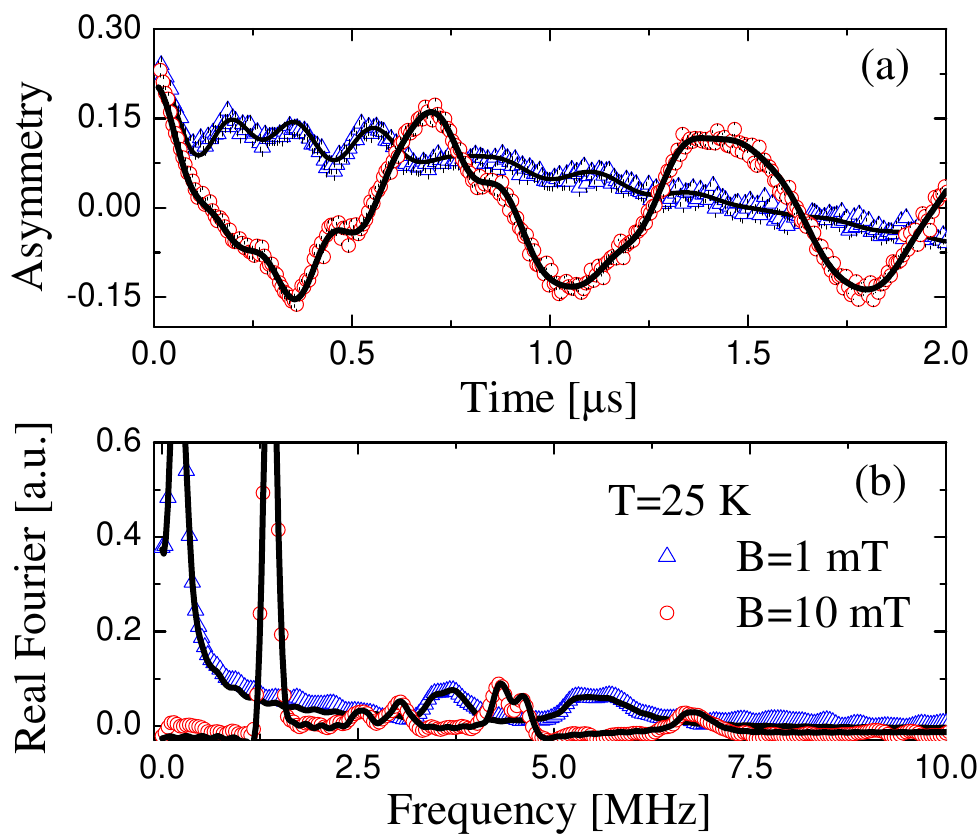}
  \caption{(Color online) (a) The asymmetries at $T=25$ K as a
    function of time for applied fields 1 and 10 mT ($B \perp
    \langle100\rangle$) and (b) the corresponding real Fourier
    transform showing at least four muonium precession frequencies in
    both cases. The solid lines are fits to a sum of precessing and
    damping signals.}
  \label{AsyField}
\end{figure}

Closer investigation of the measured asymmetries in the bulk below
$70$ K reveals that the polarization contains additional components
with precession frequencies higher than $\nu_0$. Example asymmetries,
measured at $25$~K and applied fields of 1 and 10 mT are shown in
Fig.~\ref{AsyField}(a). Note that the additional frequencies are of
the order of few MHz [see Fig.~\ref{AsyField}(b)], which we attribute
to Mu precession frequencies.
These are directly related to the hyperfine interaction of Mu and
confirm its formation at these temperatures. The size of these
frequencies ($<10$~MHz) indicates that a ``shallow'' Mu state is
formed, i.e. much smaller than the $\sim 4.4$~GHz frequency observed
for vacuum Mu.

The temperature dependence of $f_{\mu^+}$ is related to the Mu
fraction, $f_{\rm Mu}=1-f_{\mu^+}$. It can be calculated following a
semi-empirical model \cite{Cox06JPCM2},
\begin{equation} \label{Activation}
f_{\mu^+} = 1 - f_{\rm Mu}= 1 - \frac{f_0}{1+N \exp (-\Delta/T)},
\end{equation}
where $f_0$ is the normalized maximum Mu fraction, $N$ is a
density-of-states parameter and $\Delta$ is the Mu activation energy
which can be considered as an effective ionization or binding energy
\cite{Cox06JPCM2}. A fit of $f_{\mu^+}$ in bulk (circles in
Fig.~\ref{AsyT}), between 25 and 200 K, yields an ionization energy of
$\Delta=600 \pm 15$ K or $52 \pm 2$ meV. Similar fits for the data at
low implantation energies yield $\Delta=515 \pm 15$~K ($45 \pm 2$~meV)
and $270\pm 10$~K ($23\pm1$~meV) for 14.1~keV and 1.6~keV,
respectively. Similar decrease in shallow muonium ionization energy
near the surface has been recently detected in ZnO and CdS. This
effect is attributed to the presence of electric fields due to band
bending near the free surface \cite{Prokscha14condmat}.

It is worthwhile noting here that the absence of (or weak) energy
dependence of the diamagnetic fraction below $\sim 10$~K indicates
that the Mu formation at low $T$ does not depend on the number of
track electrons, i.e., there is no \textit{delayed} Mu formation below
$T \sim 10$ K.  Mu can form in two ways: i) \textit{promptly} during
charge-exchange collisions with subsequent thermalization as neutral
Mu, and ii), \textit{delayed}, where the positive muon thermalizes at
an interstitial lattice site, followed by a capture of an electron
from its own ionization track. The latter is significantly suppressed
if the number of track electrons is $\lesssim 10^3$, corresponding to
implantation energies of less than a few keV \cite{Prokscha07PRL}.  It
is also suppressed in materials with a large dielectric constant,
$\varepsilon$, where the electric field of the muon point charge is
effectively screened by the surrounding medium, thus suppressing the
probability of capturing a track electron. In STO, $\varepsilon \sim
300$ at room temperature, reaches $\sim 10000$ at $\sim25$~K and then
saturates at $20000-25000$ below 10 K \cite{Mueller79PRB}. The
increase of the diamagnetic fraction below $\sim 30$~K in the bulk may
be due to the increasing $\varepsilon$ at these temperatures, as
\textit{delayed} Mu formation becomes less likely. The low temperature
flattening of $f_{\mu^+}$ corresponds well to the saturation of
$\varepsilon$.

It is somewhat surprising that a shallow Mu state can be observed in a
system with such a large $\varepsilon$. For shallow donors, one
usually applies a hydrogenic effective-mass model to estimate the
binding energy of the donor to be $E_D \sim R_y
(m^*/m_e)/\varepsilon^2$, where $R_y = 13.6$~eV the Rydberg constant ,
$m_e$ the electron mass, and $m^*$ the conduction-band effective mass
of the electron \cite{Cox06JPCM2}. This approximation is justified by
the large electron wave-function spreading over several lattice sites.
This implies that at 25~K, where $\varepsilon \sim 10000$ in STO, the
binding energy should be zero, i.e. no Mu should form. Therefore, the
observation of a shallow Mu state indicates that it might have a more
localized polaronic character \cite{Cox06JPCM2}. Furthermore, it is
known that doping of STO may increase $m^*/m_e$ up to $\sim 20$
\cite{Ravichandran11PRB}, and that an electric field may lower
$\varepsilon$ \cite{Hyun01APL}.  Hence, ``doping'' STO with Mu may
cause a local deformation of the lattice with a local modification of
$m^*$ and $\varepsilon$.

Now we turn to evaluating the hyperfine interaction tensor of Mu. We
consider a general Hamiltonian for a muon (spin $\mathbf{I}$)
interacting with an electron (spin $\mathbf{S}$) with a fully
anisotropic hyperfine interaction \cite{Senba00PRA},
\begin{equation} \label{Ham}
{\cal H}=\gamma_{e}BS_{z}-\gamma_\mu BI_z+A_Z S_{Z}I_{Z}+A_Y S_{Y}I_{Y}+A_X S_{X}I_{X}.
\end{equation}
where $\gamma_{\mu}$ and $\gamma_e$ are the muon and electron
gyromagnetic ratios, $z$ is defined by the direction of $B$ and $A_i$
($i=X,Y,Z$) are the Mu hyperfine interactions along the principal
axes.
\begin{figure}[ht]
  \centering
  \includegraphics[width=1.0\columnwidth]{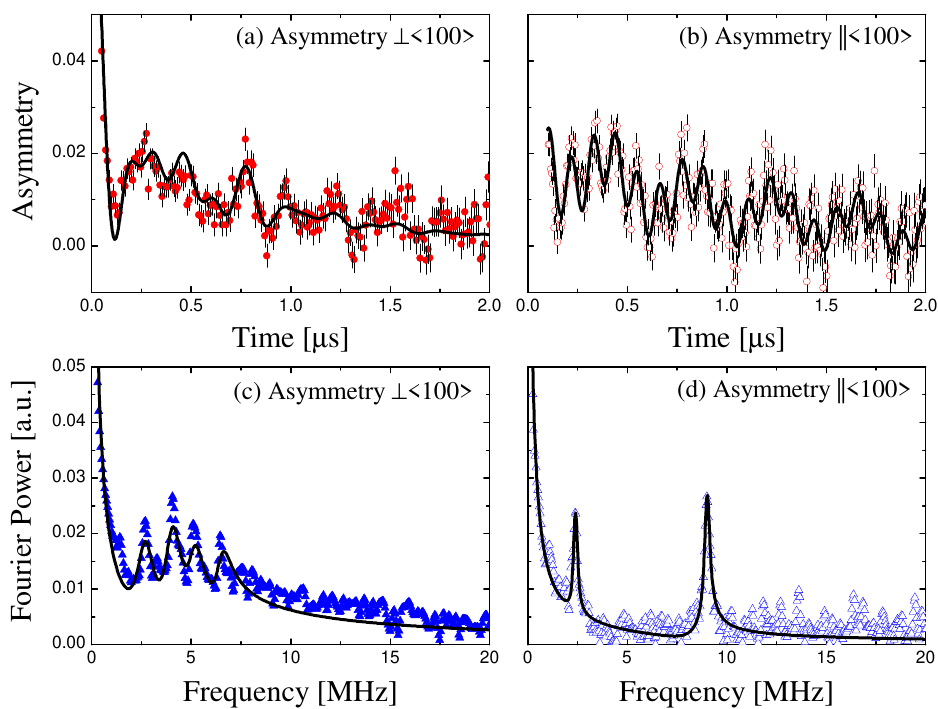}
  \caption{(Color online) The asymmetries measured (a) parallel and
    (b) perpendicular to $\langle100\rangle$ at $T=25$ K and in
    ZF. (c) and (d) are the corresponding Fourier transform showing
    clear ZF muonium precession frequencies in both cases. The solid
    lines are fits to a sum of precessing and damping signals}
  \label{ZFAsy}
\end{figure}
The coordinates $(X,Y,Z)$/$(x,y,z)$ denote the hyperfine
interaction/laboratory frame, such that the components of a spin
angular momentum vector seen in the $(X,Y,Z)$ system are expressed in
terms of the components of the same vector seen in the $(x,y,z)$
system as
\begin{equation}
\left[ \begin{array}{l} S_X \\ S_Y \\ S_Z \end{array} \right] =
D(\alpha,\beta,\gamma) 
\left[ \begin{array}{l} S_x \\ S_y \\ S_z \end{array} \right], 
\end{equation}
where $\alpha$, $\beta$,and $\gamma$ are the Euler angles of the three
consecutive rotations around $Z$, $Y$, and $Z$ axes of the $(X,Y,Z)$
coordinate system, which initially coincides with the $(x,y,z)$
system. In ZF, the Hamiltonian (\ref{Ham}) can be diagonalized
analytically to calculate the Mu frequencies, giving a maximum of 6
possible frequencies depending on the values of $A_i$
\cite{Senba00PRA}. These are the sums and differences of the different
$A_i$ parameters. Indeed our ZF measurement, shown in
Fig.~\ref{ZFAsy}(a) and (b), with different relative orientations
between the implanted muon spin and STO crystal give $\nu=2.4, 2.7,
4.1, 5.1, 6.5$ and $9$ MHz [Fig.~\ref{ZFAsy}(c) and (d)]. From these
we estimate the hyperfine parameters $A_X=1.4 \pm 0.1$ MHz,
$A_Y=6.7\pm 0.1$ MHz and $A_Z=11.5 \pm 0.1$ MHz.

\begin{figure}[ht]
  \centering
  \includegraphics[width=0.8\columnwidth]{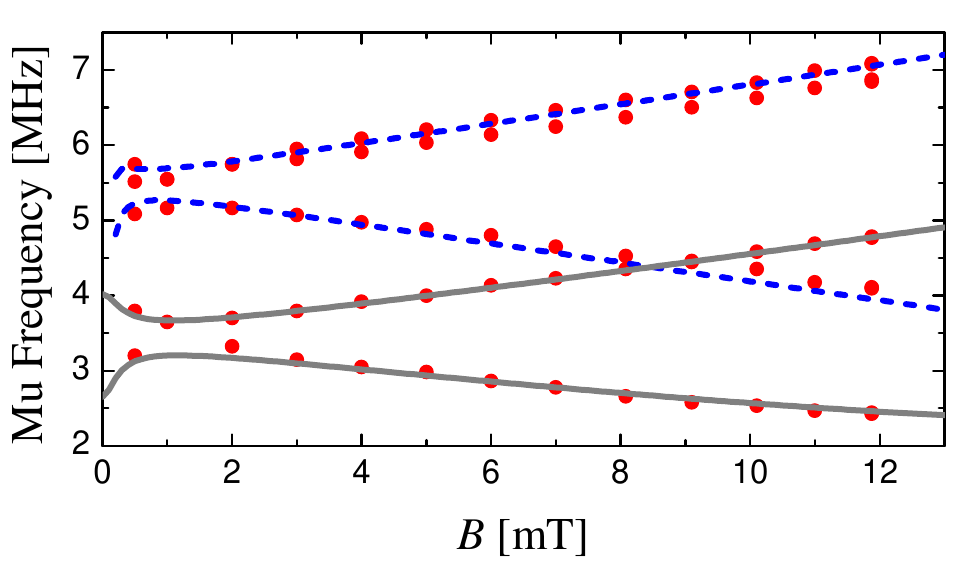}
  \caption{(Color online) The observed Muonium precession frequencies
    as a function of applied field (with an angle of $\sim 14^\circ$
    between $B$ and $\langle010\rangle$) at $T=25$K. The lines are
    calculated values with $A_X=1.37$, $A_Y=6.67$ and $A_Z=11.52$ MHz
    with $(\alpha,\beta,\gamma)=(0^\circ,57^\circ,27^\circ)$ (solid)
    and $(0^\circ,17^\circ,0^\circ)$ (dashed).}
  \label{FrqB}
\end{figure}

Next, we extract the field dependent Mu precession frequencies from
the asymmetries measured in bulk STO at 25~K. Here, we limit ourselves
to four/five most visible frequencies for each field, as plotted in
Fig.~\ref{FrqB}.
Using $A_i$ values we calculate the Mu frequencies in an applied field
by numerical diagonalization of ${\cal H}$. The amplitudes or
probabilities of precession between different Mu energy states depend
on the initial spin direction of the muon relative to the hyperfine
principal axes. We find that in order to best model the field
dependence in Fig.~\ref{FrqB}, one has to set the
$(\alpha,\beta,\gamma)=(0^\circ,57^\circ,27^\circ)$ (solid line in
Fig.~\ref{FrqB}) and $(0^\circ,17^\circ,0^\circ \pm 45^\circ)$ (dashed
line in Fig.~\ref{FrqB}). The agreement with our experimental results
is excellent considering the uncertainties and limited resolution in
the experimental data. Note, there is a small frequency splitting in
the highest frequency branch. This may be attributed to a small amount
of mosaicity or the structural distortion in STO at this temperature
\cite{Cowley96PTSL}. Such effect results in different domain
orientations and thus a small variation in the local environment of
Mu, which can only be resolved at high frequencies. Using the
extracted hyperfine parameter to calculate the precessing amplitudes
reveals that there is no missing fraction, i.e., $f_{\mu^+} + f_{\rm
  Mu}^{\rm S} = 1$, where $f_{\rm Mu}^{\rm S}$ is the fraction of
shallow Mu. This excludes the existence of any Mu state deep in the
band gap. Although we cannot exclude that the observed neutral muonium
centre is metastable, the fact that the summ of the Mu$^+$ and Mu$^0$
fractions accounts for the full muon polarization, suggests an
equilibrium balance between the two charge states. The observed
hyperfine interaction strength and ionization energy imply a shallow
donor state, modelling the analogous hydrogen state \cite{Cox09RPP}.

Note that the angle between the STO cubic axes and the normal to the
face of the oxygen octahedron are $\sim 54^\circ$. Therefore, the
hyperfine tensor and the first set of angles (solid lines in
Fig.~\ref{FrqB}) could be attributed to a Mu occupying an interstitial
site between two O--O bonds and near the face of the oxygen octahedron
in the STO crystal. The determination of the second set of angles
(dashed lines in Fig.~\ref{FrqB}) is much less reliable since it shows
a much weaker angular dependence. Nevertheless, if we assume that the
angle between the applied field and the $\langle 010 \rangle$ is $\sim
14^\circ$, then these angles are also consistent with the same Mu
site. Our results are consistent with neutron diffraction results
\cite{Sata96PRB}, infrared absorption experiments and theoretical
studies on hydrogen defect vibrational modes
\cite{Tarun11JAP,Thienprasert12PRB} as well as other theoretical work
\cite{Varley14PRB}. However, they disagree with
Refs.~\cite{Weber86PRB,Houde87PRB,Klauer92PRL,Villamagua07PS} which
place the hydrogen on O--O bond or the face of the cube between corner
sharing Sr atoms and the O atoms at the face center. Note also that
such sites, which have high symmetry, will result in an axially
symmetric Mu hyperfine tensor. Surprisingly, we also find that the
implanted muons occupy a different site from that occupied by other
implanted impurities in STO such as Li \cite{Salman04PRB,Salman06PRL}.

In conclusion, we present a direct spectroscopic observation of a
shallow hydrogen-like muonium state in STO. This confirms a
theoretical prediction that interstitial hydrogen may act as a shallow
donor in STO \cite{Xiong07JAP,Peacock03APL,Varley14PRB}. The formation
of this muonium state appears below $\sim 70$~K and implies an
activation energy of $\sim 50$~meV in bulk which decrease to $\sim
25$~meV near the surface of the crystal. We find that the shallow
impurity state has a fully anisotropic hyperfine tensor, with $A_X=1.4
\pm 0.1$ MHz, $A_Y=6.7\pm 0.1$ MHz and $A_Z=11.5 \pm 0.1$ MHz. These
results provide strong evidence of the sensitivity of the electronic
properties of STO, and in particular its surface region, to
impurities. Finally, since hydrogen is an ubiquitous impurity, these
findings may prove crucial for interpretation of the variety of
observed phenomena at LAO/STO interfaces
\cite{Ohtomo04N,Thiel06S,Huijben06NM}. We believe that hydrogen doping
effect may be a possible explanation for the excess charge carriers at
the interfaces of LAO/STO, and therefore require a more detailed
experimental and theoretical consideration. We propose that a
systematic study of the transport properties of LAO/STO interfaces as
a function of hydrogen doping may provide quantitative information
about this effect.

\begin{acknowledgments}
  This work was performed at the Swiss Muon Source (S$\mu$S), Paul
  Scherrer Institute (PSI, Switzerland). We would like to thank Rob
  Kiefl, Kim Chow and Jo\~ao Campos Gil for useful suggestions and
  fruitful discussions. We are also grateful to Ekaterina Pomjakushina
  for her assistance with the allignment of the crystals.
\end{acknowledgments}

\bibliographystyle{apsrev}
\newcommand{\noopsort}[1]{} \newcommand{\printfirst}[2]{#1}
  \newcommand{\singleletter}[1]{#1} \newcommand{\switchargs}[2]{#2#1}

\end{document}